\documentclass[%
 reprint,
superscriptaddress,
reprint,
 amsmath,amssymb,
 aps,
prb,
floatfix,
dvipsnames
]{revtex4-2}

\usepackage{graphicx}
\usepackage{dcolumn}
\usepackage{bm}
\usepackage{hyperref}
\usepackage{tikz}
\usetikzlibrary{arrows}
\usetikzlibrary{calc}
\usepackage{enumerate}
\usepackage{caption}
\usepackage{ragged2e}
\usetikzlibrary{3d, decorations.pathreplacing}

\usepackage{subcaption}
\DeclareMathAlphabet{\mathpzc}{OT1}{pzc}{m}{it}

\usepackage[export]{adjustbox}

\usepackage{xcolor}
\usepackage{soul,ulem}
\newcommand{\p}{\partial}

\newcommand{\cE} {\mathcal{E}}

\newcommand{\sqg}{\sqrt{g}}

\usepackage{upgreek}

\newcommand{\id} {\mathbb{I}}

\definecolor{maroon}{cmyk}{0,0.87,0.48,0.32}
\definecolor{Gray}{gray}{0.85}
\definecolor{LightCyan}{rgb}{0.88,1,1}
\definecolor{babyblueeyes}{rgb}{0.7, 0.84, 0.96} 
\definecolor{bluegray}{rgb}{0.4, 0.6, 0.9}
\definecolor{bluegraym}{rgb}{0.4, 0.6, 0.99}
\definecolor{airforceblue}{rgb}{0.0, 0.53, 0.74}

\begin{document}

\title{{Geometry Dynamics in Chiral Superfluids}}

\author{Yuting Bai}
\affiliation{Department of Physics,
University of Illinois at Urbana-Champaign, Illinois 61801, USA}
\author{Gabriel Cardoso}
\affiliation{Tsung-Dao Lee Institute,
Shanghai Jiao Tong University, Shanghai, 212, China}
\author{Rajae Malek}
\affiliation{Tsung-Dao Lee Institute,
Shanghai Jiao Tong University, Shanghai, 212, China}
\author{Qing-Dong Jiang}
\email{qingdong.jiang@sjtu.edu.cn}
\affiliation{Tsung-Dao Lee Institute,
Shanghai Jiao Tong University, Shanghai, 212, China}
\affiliation{School of Physics and Astronomy, Shanghai Jiao Tong University, Shanghai 200240, China}
\affiliation{Shanghai Branch, Hefei National Laboratory, Shanghai 201315, China}

\date{\today}

\begin{abstract} 
We investigate the geometric response of chiral superfluids when coupled to a dynamic background geometry. We find that geometry fluctuations, represented by the flexural mode, interact with the superfluid phase fluctuations (the Goldstone mode). Starting from a minimally coupled theory, we derive the equilibrium conditions for a static background defined by supercurrent, curvature, and tension, and then obtain linearized equations for the propagation of the Goldstone and flexural modes. The equations reveal distinctive chirality-dependent effects in the propagation of the flexural mode. Specifically, a background supercurrent induces a chiral drag effect, localizing flexural waves at the superfluid boundary, while background curvature introduces anisotropic corrections to the superfluid phase and group velocities, as well as a tension in the flexural mode dispersion. Furthermore, curvature couples flexural and phase modes into dressed excitations, with tilted Dirac cones along the principal curvature directions. These effects provide dynamical signatures of the formation of a chiral condensate, and can be tuned by manipulating the background geometry.
\end{abstract}

\maketitle

\section{Introduction}

Geometric phases lead to surprising effects, from the precession of Foucault's pendulum to the Aharonov-Bohm phase in the motion of charged quantum particles \cite{shapere1989geometric}. In chiral phases, the appropriate definition of the theory on a general geometry often requires a definition of parallel transport, which leads to a minimal coupling of the order parameter to the background geometry. In chiral superfluids, this coupling can be simply understood as due to the covariant definition of the angular momentum of Cooper pairs. While the charged case of chiral superconductors is of great interest for potential applications, we note that chiral superfluids have been experimentally realized \cite{PhysRevLett.109.215301,PhysRevLett.90.053201}.

In chiral superfluids and superconductors, the coupling to geometry appears clearly when placing the condensate on a curved background \cite{moroz2015effective}. Gaussian curvature leads to a geometric Meissner effect \cite{PhysRevLett.120.217002}, to an effective potential which attracts vortices \cite{jiang2022geometric}, and to anomalous corrections to the mass and spin current \cite{jiang2020geometric}. Changing the background topology can change the ground state of the superfluid \cite{Volovik_1999}. On the sphere, the different value of the integrated Gaussian curvature implies a different ground state altogether, with either vortices or a domain wall between different chirality domains \cite{Moroz_2016}.

The effect of a dynamical or fluctuating geometry is much less explored. Recently, it was shown that statistical fluctuations of the background geometry can modify the vortex interactions, effectively lowering the BKT transition temperature of the superfluid \cite{cardoso2024geometry}. In this work, we investigate the case of a dynamical background geometry. Namely, we assume that the background has finite bending rigidity and study the coupling between propagating shape fluctuations - the flexural mode - and the propagating phase fluctuations of the chiral superfluid - the Goldstone mode. Starting from the minimally coupled theory, we derive the equilibrium equation to be satisfied by the background configuration, and the linearized equations of motion for phase and flexural waves propagating on top of that background. 

\begin{figure*}[!htbp]
\minipage{0.33\textwidth}
\begin{tikzpicture}
    \node[inner sep=0pt] (A){\includegraphics[height=1.2in, width=2.3in]{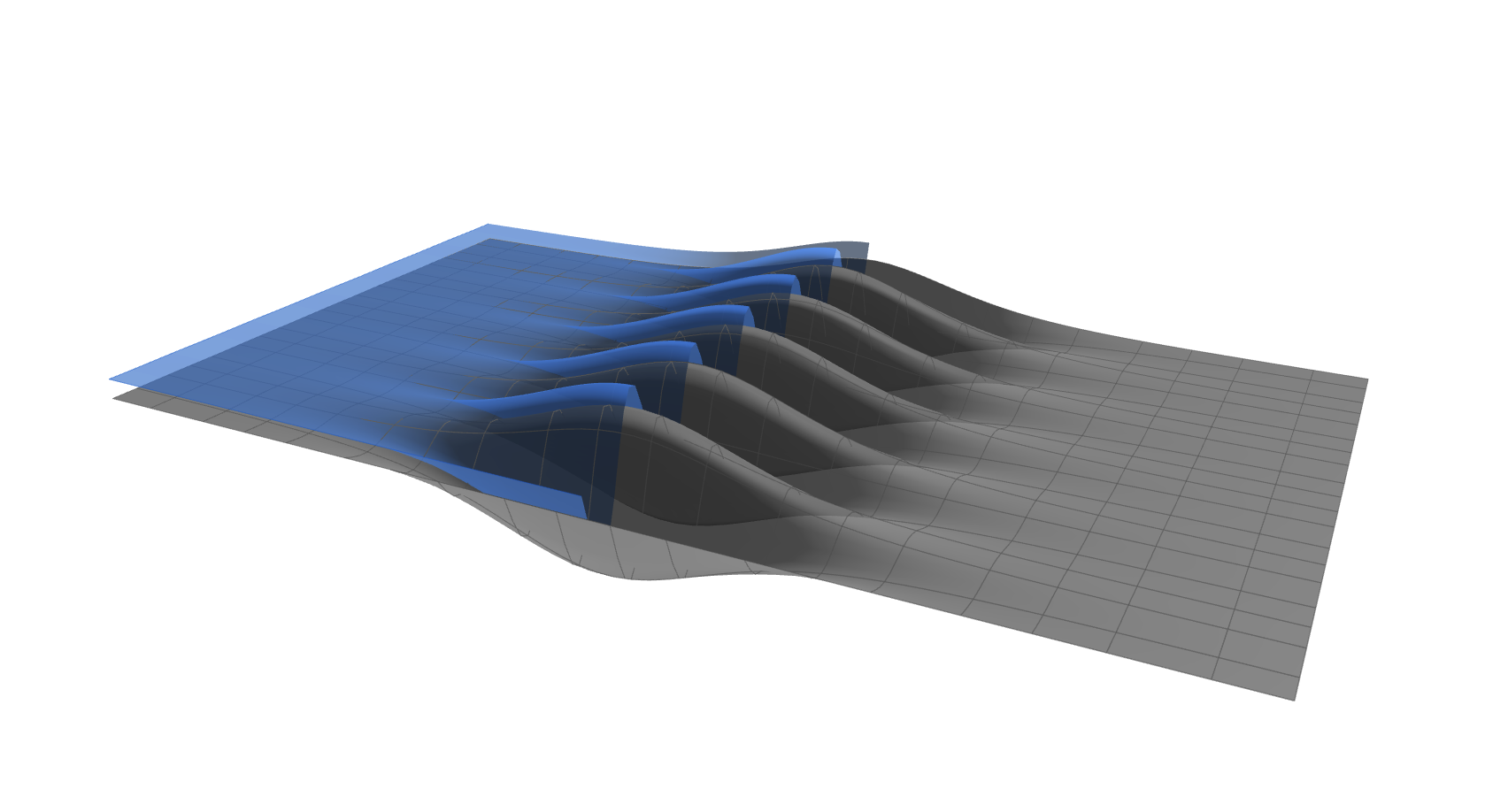}};
  
    \draw[->, thick, white] (-0.25,0) -- (0.35,-0.13) node[above] {\, $ \vec{n}$}; 
    \draw[->, thick, white] (-0.8,0.13) -- (-0.5,0.4) node[left] {\, $ \vec{J_{0}}$}; 
  \draw[thick,  ->, orange] (-1.8,0.15) arc[start angle=180, end angle=450, radius=0.18cm, y radius=0.1cm] node[right]  {\,  $\ell$};
\end{tikzpicture}
   \subcaption{} \label{fig:backgroundsupercurrent}
\endminipage  
\minipage{0.33\textwidth}
\begin{tikzpicture}
    \node[inner sep=0pt] (A){\includegraphics[height=1.2in, width=2.3in]{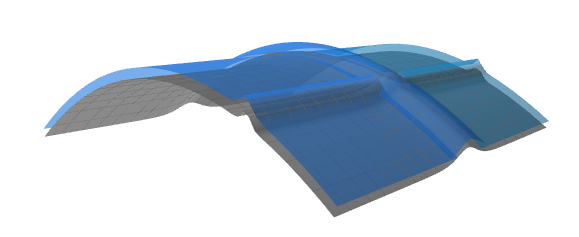}};

\draw[->, thick, Gray] (0.2,0.6) -- (-0.2,0.52) node[above] {\,   $ \delta \tilde{h}$};
\draw[->, thick, Gray] (0.4,0.2) -- (0.7,-0.2) node[left] {\,  $ \delta \tilde{h}$};
\node[text=white] at (2,0.2) { $ H_0$};

\end{tikzpicture}
\subcaption{ }
\label{fig:parabola}
\endminipage  
\minipage{0.33\textwidth}
\begin{tikzpicture}
      \node[anchor=south west,inner sep=0] (image) at (0,0) {\includegraphics[height=1.2in, width=2.3in]{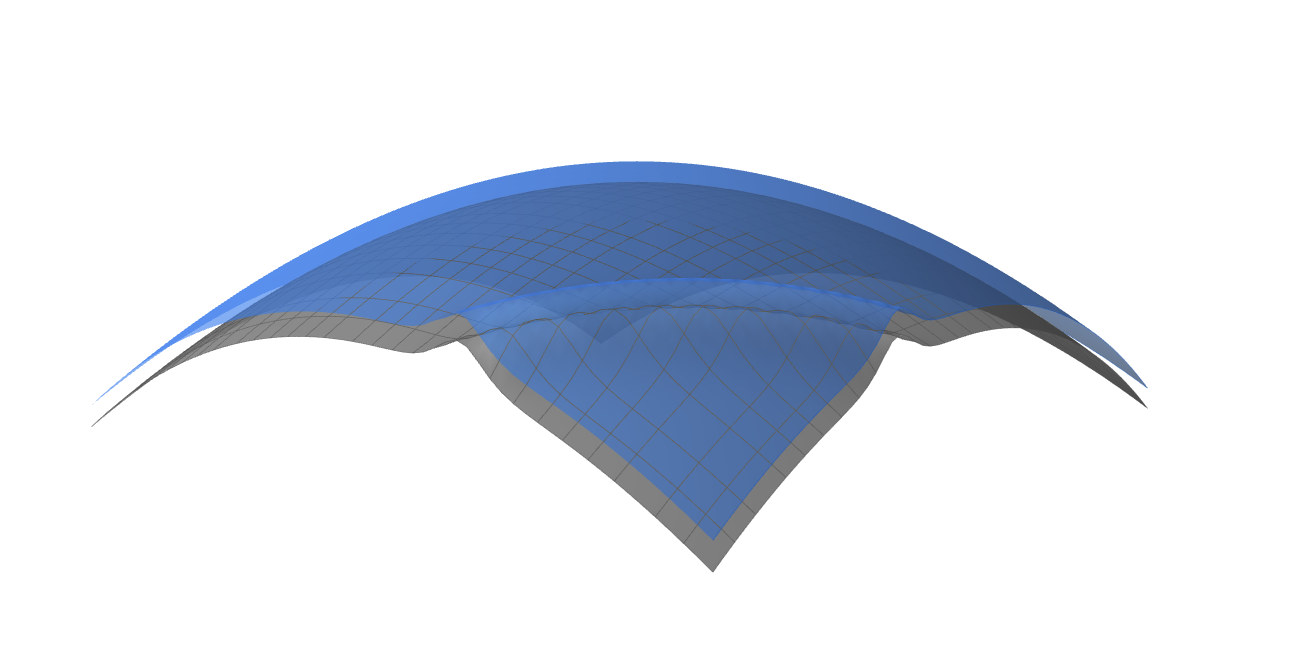}};
      \draw[->, thick, white] (0.7,0.4) -- (0.45,0.6);
        \begin{scope}[shift={(image.south west)}, x={(image.south east)}, y={(image.north west)}]
        \clip (0,0) rectangle (1,1);

\draw[->, thick, white] (0.485,0.68) -- (0.525,0.69);
\node[text=white] at (0.57,0.64) { $ \vec{J_{0}}$};
\node[text=white] at (0.35,0.62) { $ H_0, K_0$};

        \foreach \x/\y/\dx/\dy in {  
         0.43/0.73/0.043/-0.045, 0.53/0.685/0.043/0.044} {
            \draw[->,white,thick] (\x,\y) -- +(\dx,\dy);
        }
\draw[->, thick, Gray] (0.53,0.5) -- (0.53,0.35) node[right] { $ \delta \tilde{h}$};
     
    \end{scope}

\end{tikzpicture}
\subcaption{ }
\label{fig:quadratic}
\endminipage 

 \caption{ \justifying Propagation of the Goldstone and flexural modes on different background configurations. (a) A constant background supercurrent $\vec{J}_0$ changes only the boundary conditions for the flexural mode. If the superfluid occupies a subregion of the substrate and the cross product of $\vec{J}_0$ with the direction normal to the boundary $\vec{n}$ is parallel to the condensate angular momentum $\vec{\ell}\hbar$, then the superfluid can drag a boundary flexural mode. (b) A uniaxial deformation of the substrate generates a background mean curvature $H_0$ but no Gaussian curvature. This couples the Goldstone and flexural waves leading to dressed degrees of freedom, and the dressed flexural waves $\delta\tilde{h}$ feel an effective tension when propagating on the direction transverse to the background deformation axis. (c) A biaxial deformation of the substrate leads to a background with both mean curvature $H_0$ and Gaussian curvature $K_0$. In this case, the background also needs to have a background supercurrent $\vec{J}_0$ and tension $\sigma_0$. The effective tension on the spectrum of dressed flexural waves becomes isotropic in the limit of maximal Gaussian curvature $K_0\to H_0^2$.}
\label{fig:main}
\end{figure*}
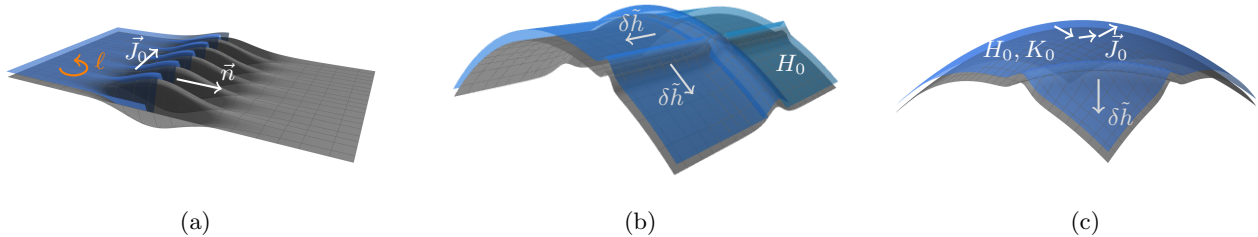

We study the three cases outlined in figure \ref{fig:main}: when the background includes a constant supercurrent $\vec{J}_0$ (Fig. \ref{fig:backgroundsupercurrent}); under uniaxial strain of the background, which introduces a mean curvature $H_0$ (Fig. \ref{fig:parabola}); and under biaxial strain, in which case the background has not only mean curvature but also Gaussian curvature $K_0$ (Fig. \ref{fig:quadratic}). We find that in each case the coupling modifies the propagation of the flexural and Goldstone modes, such as localizing flexural waves at the boundary of the superfluid, hybridizing the flexural and Goldstone modes into polariton-like dressed modes, and generating a tension on the direction transverse to the curvature. These effects can play a role in the physics of two-dimensional materials, where the coupling to the highly fluctuating flexural deformations has been shown to modify electronic transport properties at low temperatures \cite{PhysRevLett.100.076801,PhysRevB.88.115418,PhysRevB.83.174104,doi:10.1126/sciadv.abn8007,lopez2022effect}, and can serve as a signature of the formation of a chiral condensate.

The paper is organized as follows. In section \ref{sec:Modes}, we review the bare Goldstone and flexural modes, as well as their minimal coupling in the Ginzburg–Landau theory. In section \ref{sec:background}, we derive the equilibrium equation for the background and the linearized equations for propagating waves. In sections \ref{sec:backgroundcurrent}, \ref{sec:H0} and \ref{eq:K0} we discuss the effects of background current, extrinsic curvature, and Gaussian curvature, respectively. We present our main conclusions and possible future directions in section \ref{sec:conclusion}. Appendices are included for more details on the calculations.

\section{Goldstone and Flexural modes}\label{sec:Modes}

We consider the minimal model of a chiral superfluid on a curved surface, given by the Ginzburg-Landau action,
\begin{align}
    \int dt d^2r \sqg\left [ i\hbar \psi ^*D_t\psi- \frac{\hbar ^2}{2m} g^{ij}D_{i}\psi^* D_{j}\psi -V(\vert \psi \rvert)   \right],\nonumber
\end{align}
where $D_\mu=\partial_\mu+\ell\Omega _\mu$ is the covariant derivative on the surface and the integer $\ell$ labels the angular momentum $\ell\hbar$ of the Cooper pairs. In terms of local vielbein $\hat{e}_{1,2}$, the spin connection has the form $\Omega_\mu = \hat{e}_1 \cdot \partial_\mu \hat{e}_2$. The potential $V(|\psi|)$ gives a finite expectation value for the superfluid density $|\psi|$ so that, integrating out density fluctuations, leaves
\begin{align}
    S_\theta &=\frac{1}{2}\int dt d^2r\sqrt{g}[\gamma_0 (\partial _0\theta  +\ell\Omega_0)^2\nonumber \\
    &\hspace{2.5cm}-\gamma g^{ij} (\p_i\theta +\ell\Omega_i)(\p_j\theta +\ell\Omega_j)],\label{eq:thetaaction}
\end{align}
with the effective superfluid stiffness $\gamma$ fixed by the bare mass $m$ and the mean superfluid density. In the fixed planar geometry $g_{ij}=\delta_{ij}$, this action describes the propagation of phase fluctuations known as the superfluid Goldstone mode, with massless dispersion
\begin{align}
    &\omega(k) =  u\lvert k \rvert , &u^2 = \frac{\gamma}{\gamma_0}. \label{eq:Goldstoneomega}
\end{align}

For a dynamical background, one should add an action for the geometry. In our setup, we consider the case of a flexible two-dimensional membrane. Its dynamics is controlled by elastic forces which, at large length scales, lead to the resistance against curvature and bending \cite{helfrich1973elastic,bowick2001statistical,nelson2004statistical}. The leading contributions are given by the potential energy density terms

\begin{align}
v = \sigma + \frac{\kappa _r}{2}H^2 + \kappa_G  K,
\end{align}
where $H$ and $K$ are the local values of the mean curvature and the Gaussian curvature, $\sigma$ is the surface tension, $\kappa_r$ is the bending rigidity, and $\kappa_G$ is the Gaussian rigidity. In a crystalline membrane, $\kappa_r$ arises from the gradient expansion of the bending energy to the leading order \cite{PhysRevA.38.1005}. The contribution of the Gaussian curvature term is topological and does not contribute to the equations of motion. We are interested in local effects, so that it is natural to take smooth shapes parametrized as $\vec{r}(t)=(x,y, h(t,x,y))$, where the height function $h(t,x,y)$ is smooth. Expanding the total action gives:
\begin{align}
    \frac{1}{2}\int dt d^2r\Big[\gamma_0 J_\mu J^{\mu}+\kappa_0(\partial_t h)^2-\sigma (\nabla h)^2-\kappa_r(\Delta h)^2
    \Big],\label{eq:totalaction}
\end{align}
where we also included the kinetic energy of the membrane, with mass density $\kappa_0$. Here, we introduced the notation $J_{\mu} = \p_\mu\theta+\frac{\ell}{2}\epsilon ^{jk}\partial _{k}[(\partial _{j}h)(\partial _{\mu }h)]$ for the lowest-order expansion of the supercurrent, and the summation over repeated greek indices $a^{\mu}a_{\mu} = (a_0)^2-u^2(\vec{a})^2$. Note that, in our approximation, we match the orders in derivatives of $h$ and $\theta$ so as to expand in the covariant supercurrent $J_\mu$.

Besides the Goldstone mode, the quadratic part of the action also leads to the propagation of fluctuations of the membrane shape $h$, with the dispersion 
\begin{align}
    \omega^2 =&\frac{k^4}{4m_\kappa^2}+\sigma k^2, &m_\kappa^2=\frac{\kappa_0}{4\kappa_r}.\label{eq:flexuralomega}
\end{align}
In the absence of tension, this gives the massive dispersion relation $\omega = \pm\frac{k^2}{2m_\kappa}$ while, in the presence of tension, $\omega\sim\pm\sigma k$ becomes linear at small wavevectors. This mode is known as the flexural mode of the membrane, and it plays a role in the physics of two-dimensional materials. Since the covariant current depends on the height field $h$ through the definition of the spin connection, the crossed term in the action \eqref{eq:totalaction} minimally couple these two modes. This coupling modifies both the equilibrium configuration as well as the propagating modes, as we now consider.

\section{Equations of motion on a fixed background}\label{sec:background}

Note that at this order of approximation the coupling between the $\theta$ and $h$ fields gives a cubic term in the total action (\ref{eq:totalaction}). One approach is to expand on this coupling and perturbatively calculate the renormalization of the superfluid and membrane properties \cite{cardoso2024geometry}. Alternatively, we consider the mean-field effects of a background supercurrent and geometry profile, by expanding
\begin{align}
    &h = h_0(x,y)+\delta h(x,y,t), &\theta = \theta_0(x,y)+\delta\theta(x,y,t),\nonumber
\end{align}
where the background values $h_0$, $\theta_0$ are a static saddle-point of the action.

The second-order expansion of (\ref{eq:totalaction}) gives

\begin{align*}
S_0[h_0, \theta_0] &+ \int K_{\alpha}^{(1)}[h_0, \theta_0](r, t) \delta \phi^{\alpha}(r, t) \, dr  \nonumber \\
& + \int \int' K_{\alpha \beta}^{(2)}[h_0, \theta_0](r, t; r', t') \delta \phi^{\alpha}(r, t) \delta \phi^{\beta}(r', t'),
\end{align*}
where $\delta \phi^{1,2}$ denote the fields $\delta h$, $\delta\theta$. The saddle point equation $K^{(1)}[h_0,\theta] = 0$ corresponds to the equilibrium of the background configuration,

\begin{align}
&\partial_{\mu} J_0^{\mu} = 0, \label{eq:j0background} \\
&\kappa_0 \partial_t^2 h_0 - \partial_i \left( \sigma \partial_i h_0 \right) - \kappa_r \Delta^2 h_0 = \nonumber \\
&\hspace{1cm} \frac{\gamma_0 \ell}{2} \epsilon^{ij} \left[ \partial_j \partial_{\mu} \left( J_0^{\mu} \partial_i h_0 \right) - \partial_i \left( J_0^{\mu} \partial_j \partial_{\mu} h_0 \right) \right]. \label{eq:h0background}
\end{align}
Equation (\ref{eq:j0background}) is the continuity equation for the background supercurrent, and (\ref{eq:h0background}) is the equation for the mechanical equilibrium of the background membrane shape. Alternatively, equation (\ref{eq:h0background}) determines the background tension distribution $\sigma(r)$ which should be applied to pin the membrane to the shape $h_0(r)$, as we will consider for the different background configurations. Given a solution of (\ref{eq:j0background},\ref{eq:h0background}), the spectrum of shape fluctuations is determined by the quadratic part of the action,
\begin{align}
&\delta S^{(2)}=\delta S_{\theta \theta}+ \delta S_{\theta h} + \delta S_{hh}, \label{eq:QuadraticAction} \\ 
&\delta S_{\theta \theta }=\int dtd^2r\frac{\gamma_0}{2}\partial ^{\mu}\delta \theta \partial _{\mu}\delta \theta,\nonumber\\
&\delta S_{\theta h}=\int dtd^2r\frac{\gamma_0 \ell}{2}\epsilon ^{ij}\partial ^{\mu}\delta \theta (\partial _ih_0\partial _j\partial_{\mu}\delta h+\partial _i\delta h\partial _j\partial _{\mu}h_0),\nonumber\\
&\delta S_{hh}=\int dtd^2r\left [ \frac{\kappa _0}{2}(\partial _t \delta h)^2-\frac{\sigma _0}{2}(\nabla \delta h)^2-\frac{\kappa _r}{2}(\Delta\delta h)^2    \right ] \nonumber\\
&\hspace{.5cm}+\int dtd^2r \frac{\gamma_0 \ell}{2}J_0^{\mu} \partial _j(\epsilon ^{ij}\partial _i\delta h \partial _{\mu}\delta h) \nonumber\\
&\hspace{.5cm}+\int dtd^2r\frac{\gamma_0}{2} \left[\frac{\ell}{2}\epsilon^{ij}(\partial_i h_0 \partial_j \partial^{\mu}\delta h + \partial_i \delta h \partial_j \partial^{\mu}h_0) \right]^2,\nonumber
\end{align}
where the coefficients of the action are evaluated at the background configuration. Note that: (i) the mixed term $S_{\theta h}$ is non-zero only in a curved background; (ii) the effect of the coupling to the chiral superfluid ($\ell\neq 0$) appears not only in the mixed term $S_{\theta h}$, but also on the membrane part, through the second and third terms in $S_{hh}$; (iii) the second term in $S_{hh}$ only modifies the equations of motion when the background current is non-uniform. Otherwise, it reduces to a total derivative; (iv) the third term arises from the contact term $\Omega_{\mu}\Omega^{\mu}$, which is necessary to maintain the reparametrization invariance to this order.

The saddle-point of action \eqref{eq:QuadraticAction} gives the linearized equations for the Goldstone and flexural modes, which are valid for small perturbations around the background configuration, and in the small gradient regime. The former approximation is a small amplitude approximation, while the latter is a long-wavelength approximation. We find that interchanging the order of the two approximations leads to the same equations at this order. Although the effective action \eqref{eq:QuadraticAction} for the propagating modes is quite complicated, the equations of motion can be brought to a much simpler form in interesting examples. We study the configurations listed in figure \ref{fig:main}: a constant background supercurrent on with a domain wall; a uniaxial bending which leads to mean curvature $H_0$ but not Gaussian curvature; and finally a generic curved region with nonvanishing mean curvature $H_0$ and Gaussian curvature $K_0$.

\section{Background supercurrent}\label{sec:backgroundcurrent}

We first consider the effect of a uniform background supercurrent, with no background tension, $\sigma=0$. Then flexural mode decouples from the Goldstone mode, and its effective action is given by
\begin{align}
\delta S_{hh}&=\int dtd^2r\left [ \frac{\kappa_0 }{2}(\partial _t \delta h)^2 - \frac{\kappa _r}{2}(\Delta\delta h)^2    \right ] \nonumber \\
&\hspace{1.5cm}+\int dtd^2r \frac{\gamma_0 \ell}{2}J_0^{\mu}\partial _j (\epsilon ^{ij}\partial _i \delta h \partial _{\mu}\delta h).
\label{Superfluid Junction Action}
\end{align}
If the background current $J_0^{\mu}$ is uniform, then the extra term is a total derivative and the equation of motion,
\begin{equation}
    \left[\p_t^2+\frac{\Delta^2}{4m_\kappa^2}\right]\delta h=0,
\end{equation}
is not modified. Still, this term modifies the boundary conditions. As we now discuss, these can lead to the chiral drag of boundary flexural waves and to the anomalous reflection of the flexural mode at the boundary. We investigate these effects in the following setup: two regions of different background supercurrents $\vec{J}_1$ and $\vec{J}_2$ are joined at a planar boundary at $x=0$. We take $\vec{J}_1$ and $\vec{J}_2$ constant and pointing along the direction parallel to the boundary, so that there is no superflow across the domain wall. Rather, we are interested in the implications of the modified boundary conditions on the propagation of geometric waves across the domain wall. We illustrate the setup in figure \ref{fig:backgroundsupercurrent} for the case $\vec{J}_1=\vec{J}_0$, $\vec{J}_2=0$, where it corresponds to the boundary of a finite region occupied by the superfluid.

Varying over the boundary value of $\delta h$ gives a jump condition on the third derivatives,
\begin{align}
(\kappa _r \partial _x \nabla ^2h+\gamma_0 \ell u^2J_1\partial _y^2 h)|_{+}&=(\kappa _r \partial _x \nabla ^2h+\gamma_0 \ell u^2J_2\partial _y^2 h)|_{-},\nonumber
\end{align}

where $|_\pm$ denotes the limiting values at $x\to 0^{\pm}$, while the lower-order derivatives of $\delta h$ are continuous across the domain boundary. Using the translational symmetry in the $y$ direction, we can simplify the equation of motion to
\begin{align}
\left[\p_t^2 + \frac{(\partial _x^2 - k_y^2)^2}{4m_\kappa^2}\right]\delta h=0,
\end{align}
and the jump condition to
\begin{align}
\partial_x^3 \delta h|_{+}-\partial_x^3 \delta h|_{-}=b k_y^2\delta h,\label{eq:bcond}
\end{align}
where
\begin{equation}
    b=\frac{\gamma\ell (J_1-J_2)}{\kappa_r},
\end{equation}
and we replaced $\delta h$ by its Fourier transform in $y$. While in the bulk the solutions are propagating waves $\delta h(x,t) = e^{i(\omega t-k_xx)}$ with dispersion \eqref{eq:flexuralomega}, we now show that the modified boundary condition \eqref{eq:bcond} leads to exponentially localized solutions $e^{i\omega t+\lambda x}$ at the boundary.

\subsection{Chiral drag of flexural waves}

\begin{figure}[]

    \includegraphics[width=.8\linewidth,left]{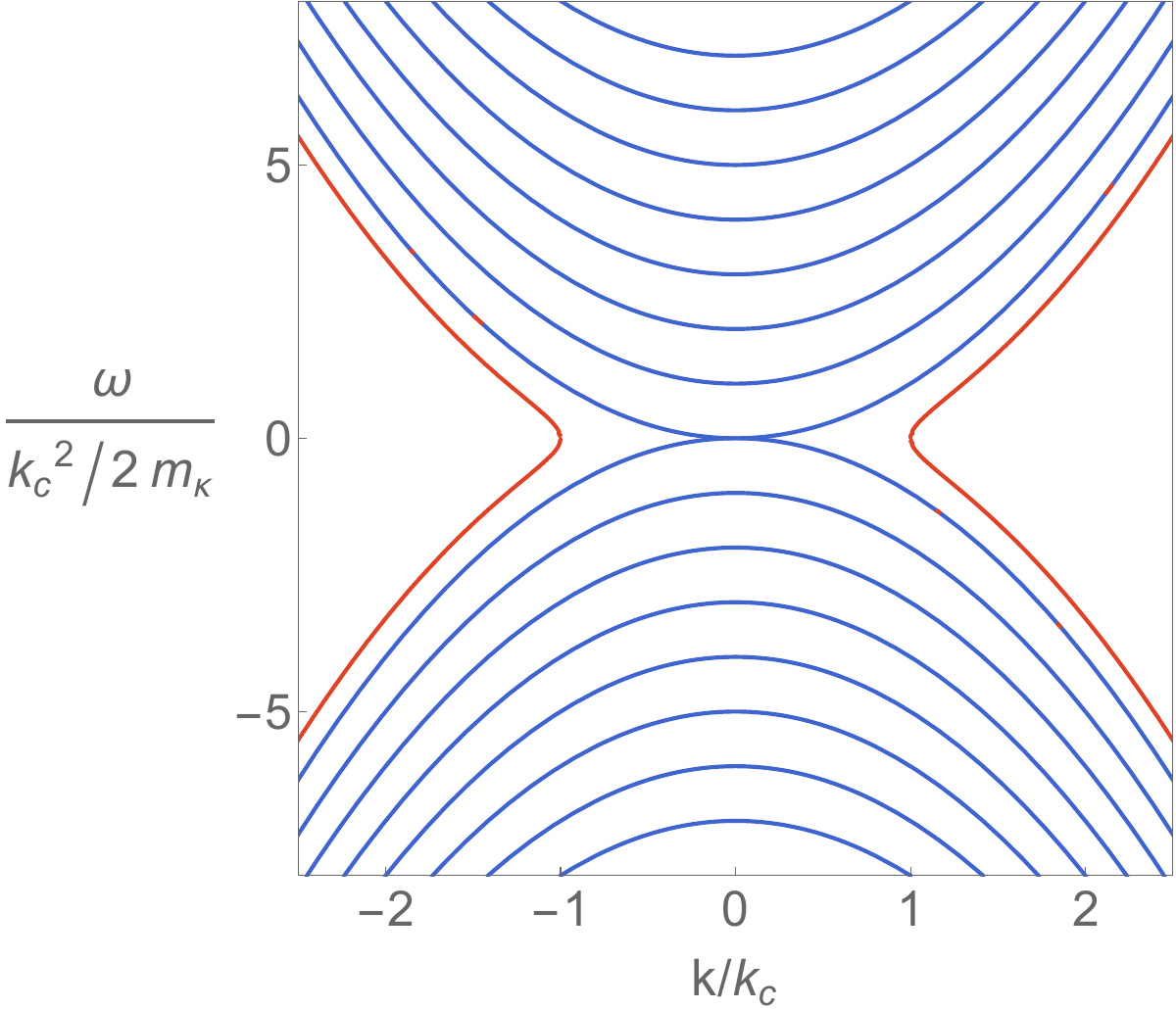}
    \caption{  \justifying Dispersion of the bulk (blue, for different values of $k_x$) and boundary (red) flexural waves as a function of the momentum parallel to the boundary, $k_y$. The boundary waves only exist for $\ell(J_1-J_2)>0$ and for $|k_y|>k_c$, equation \eqref{eq:kcJ}. We normalize the axes by $k_c$.}
    \label{fig:boundaryspectrum}
\end{figure}

Note that the secular equation,
\begin{eqnarray}
    \omega^2=\frac{(k_x^2+k_y^2)^2}{4m_\kappa^2},\label{eq:omegakxy}
\end{eqnarray}
is of fourth order in $k_x$, the momentum normal to the boundary. The nature of the solutions depends on the values of $\omega$ and $k_y$. For 
\begin{equation}
    \omega^2<\frac{k_y^4}{4m_\kappa^2},
\end{equation}
all four solutions are purely imaginary,
\begin{align}
    &k_x=\pm i\lambda_{1,2}, &\lambda_{1,2}=\sqrt{k_y^2\pm 2m_\kappa\omega}.
\end{align}
The boundary conditions far from the boundary fix the sign of the exponents, so that the solution is of the form $ e^{i(\omega t-k_yy)}\delta h(x)$, with
\begin{align}
\delta h(x)=
\left\{\begin{matrix}
C_1 e^{-\lambda _1 x}+C_2e^{-\lambda _2 x}, &\quad x>0 \\
C_3 e^{\lambda _1 x}+C_4e^{\lambda _2 x }, &\quad x<0
\end{matrix}\right..
\end{align}
As we discuss in Appendix \ref{Derivation for 1st Setup}, this solution can only satisfy the boundary conditions at $x=0$ if the coefficient $b$ in \eqref{eq:bcond} is positive. Thus we find that there are only flexural waves localized at the boundary for
\begin{align}
\left\{\begin{matrix}
J_1>J_2, &\quad \text{if }\ell>0 \\
J_1<J_2, &\quad \text{if }\ell<0
\end{matrix}\right..
\label{eq:J12condition}
\end{align}
We also find that, when present, the boundary waves have a minimal value for the momentum along the boundary,
\begin{eqnarray}
    |k_y|> k_c = \frac{\gamma \ell (J_1-J_2)}{4\kappa_r}.\label{eq:kcJ}
\end{eqnarray}
For $|k_y|\gtrsim k_c$, the one-dimensional dispersion relation for boundary waves is approximately given by
\begin{align}
\omega (k_y) \approx \pm \frac{k_y^2}{m_\kappa}\sqrt{\frac{2}{5}\left(\frac{|k_y|}{k_c}-1\right)}.\label{eq:bwavesomega}
\end{align}
This dispersion is plotted in figure \ref{fig:boundaryspectrum}, together with the bulk bands for different values of $k_x$.

We note that the condition 
\eqref{eq:J12condition} for the existence of boundary waves is chiral. Indeed, the jump $J_2-J_1$ is proportional to the integral of the curl $\nabla\times J$ across the boundary between the two domains, which in our setup is just $\partial_x J_y$ and is localized at the boundary. Equation \eqref{eq:J12condition} says that there are localized boundary flexural waves only when the value of the curl of $J$ across the boundary is opposite to the angular momentum of the superfluid $\ell\hbar$. Alternatively, one can consider the case where $J_1=J_0$, $J_2=0$, in which case the flexural mode can propagate without obstruction on the substrate but the superfluid is only present in the $x<0$ half space. Then the cross product $\vec{L}_z\times\vec{n}$ between the angular momentum of the condensate and the vector normal to the boundary of the superfluid defines an orientation along the boundary (see Fig. \ref{fig:backgroundsupercurrent}). If the supercurrent $\vec{J}_0$ flows along this direction on the boundary, then the superfluid will drag boundary flexural waves along the boundary, but not if the supercurrent flows opposite to this direction. Finally, if the region occupied by the superfluid has a strip geometry, then the flexural mode gets localized on only one of the boundaries. Thus the existence of boundary flexural waves is a chiral effect, which manifests in the properties of the flexural mode through the geometry coupling.

\subsection{Anomalous Reflection}

Another interesting effect of the geometric coupling to background supercurrent is the anomalous reflection of flexural waves. Namely, even though the equation of motion is the same on both sides of the domain wall (figure \ref{fig:backgroundsupercurrent}), matching the boundary conditions leads to a non-vanishing reflected wave, as well as a boundary wave component.

In order to have solutions propagating in the $x$ direction, we now consider the bands with
\begin{eqnarray}
    \omega^2>\frac{k_y^4}{4m_\kappa^2}.
\end{eqnarray}
Then \eqref{eq:omegakxy} has two propagating solutions $e^{i(\omega t- k_x x)}$, with
\begin{align}
    k_x = \pm\sqrt{2m_\kappa|\omega|-k_y^2}
\end{align}
and two exponential solutions $e^{i\omega t+\lambda x}$, with
\begin{align}
    &\lambda = \pm\sqrt{2m_\kappa|\omega|+k_y^2}.
\end{align}

Interestingly, a superposition of the incident, reflected and transmitted waves is not enough to solve all the boundary conditions. Instead, one needs to include a component of the localized solutions,
\begin{align}
\delta h(x)=
\left\{\begin{matrix}
e^{ik_xx}+re^{-ik_xx}+C_1e^{\lambda x}, &\quad x<0 \\
te^{ik_xx}+C_2e^{-\lambda x}, &\quad x>0
\end{matrix}\right..
\end{align}

\begin{figure}
    \centering
    \includegraphics[width=0.9\linewidth]{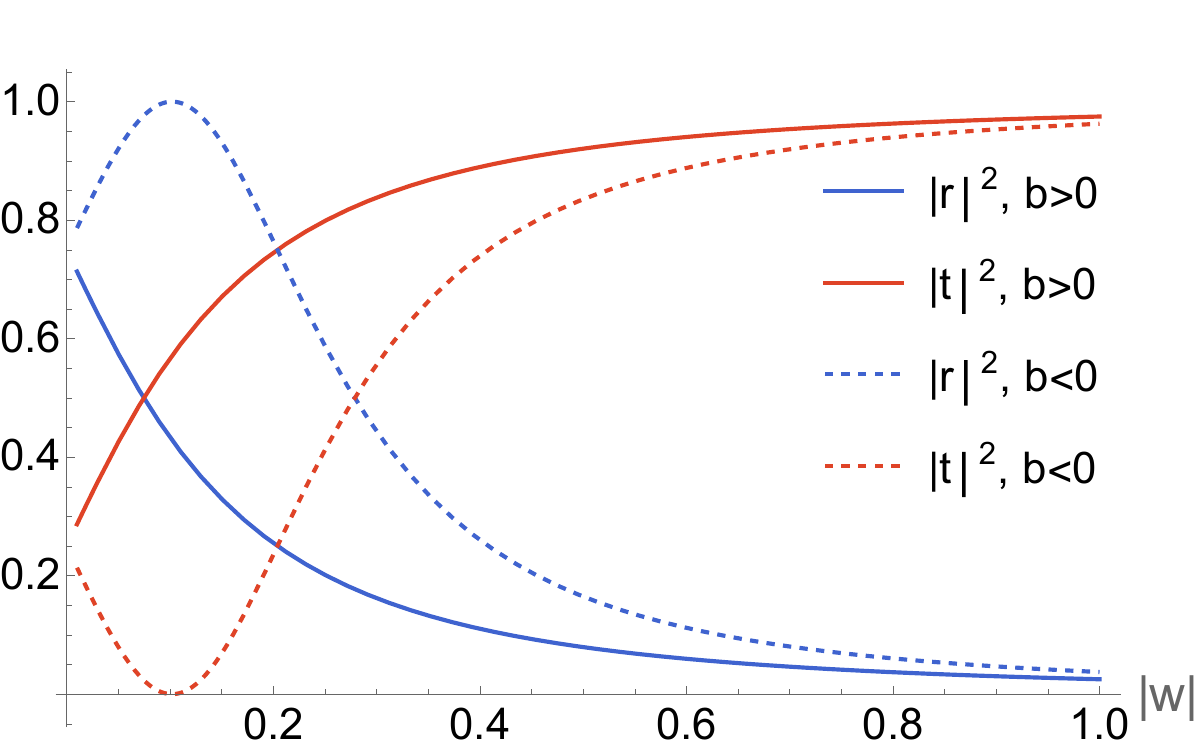}
    \caption{\justifying Frequency dependence of the reflection and transmission amplitudes of the flexural mode across the boundary between two domains with different background supercurrents $J_1$ and $J_2$. The square-root normalized frequency parameter is given by $w=\frac{\sqrt{2m_\kappa|\omega|}}{b}$, and the incidence angle is $\theta=\frac{\pi}{4}$. At small frequencies, the reflection and transition amplitudes depend on the sign of the chirality parameter $b\propto\ell (J_1-J_2)$, with perfect reflection of the waves with frequency given by \eqref{eq:omegatheta} in the case of $b<0$.}
    \label{fig:Reflection Amplitude}
\end{figure}

Solving the boundary conditions gives the reflection and transmission coefficients (see Appendix \ref{Derivation for 1st Setup})
\begin{align}
    r &= \frac{-\sin^2 \theta  \sqrt{1+\sin^2\theta} }{4iw \cos\theta\sqrt{1+\sin^2\theta }+\sin^2\theta (\sqrt{1+\sin^2\theta }+i \cos\theta  ) },\nonumber\\
    t &= \frac{ i\cos\theta (\sin^2\theta +4w\sqrt{1+\sin^2\theta }  ) }{4iw\cos\theta\sqrt{1+\sin^2\theta }+\sin^2\theta (\sqrt{1+\sin^2\theta }+i\cos\theta )},\nonumber
\end{align}
where $\theta=\arctan\left(\frac{k_y}{k_x}\right)$ is the incidence angle and we defined the square-root frequency parameter
\begin{eqnarray}
    w=\frac{\sqrt{2m_\kappa|\omega|}}{b},
\end{eqnarray}
which depends on the sign of $b\propto\ell(J_1-J_2)$. Importantly, in deriving these equations we find that the amplitudes of the localized solutions, $C_{1,2}$, are non-vanishing, so that the boundary waves also play a role in determining the reflection and transmission coefficients. Note that $(r,t)\to (0,1)$ in the limit $b\to 0$, corresponding to full transmission in the absence of a domain wall. For $b\neq 0$, we can define the phase shift $\phi$ of the reflected wave by $r = \lvert r \rvert e^{i\phi}$, and we find
\begin{align}
\tan \phi = -\frac{\cos\theta(\sin^2\theta +4w\sqrt{1+\sin^2\theta })}{\sin^2\theta \sqrt{1+\sin^2\theta }  } .
\end{align}

\begin{figure}
    \centering
    \includegraphics[width=0.9\linewidth]{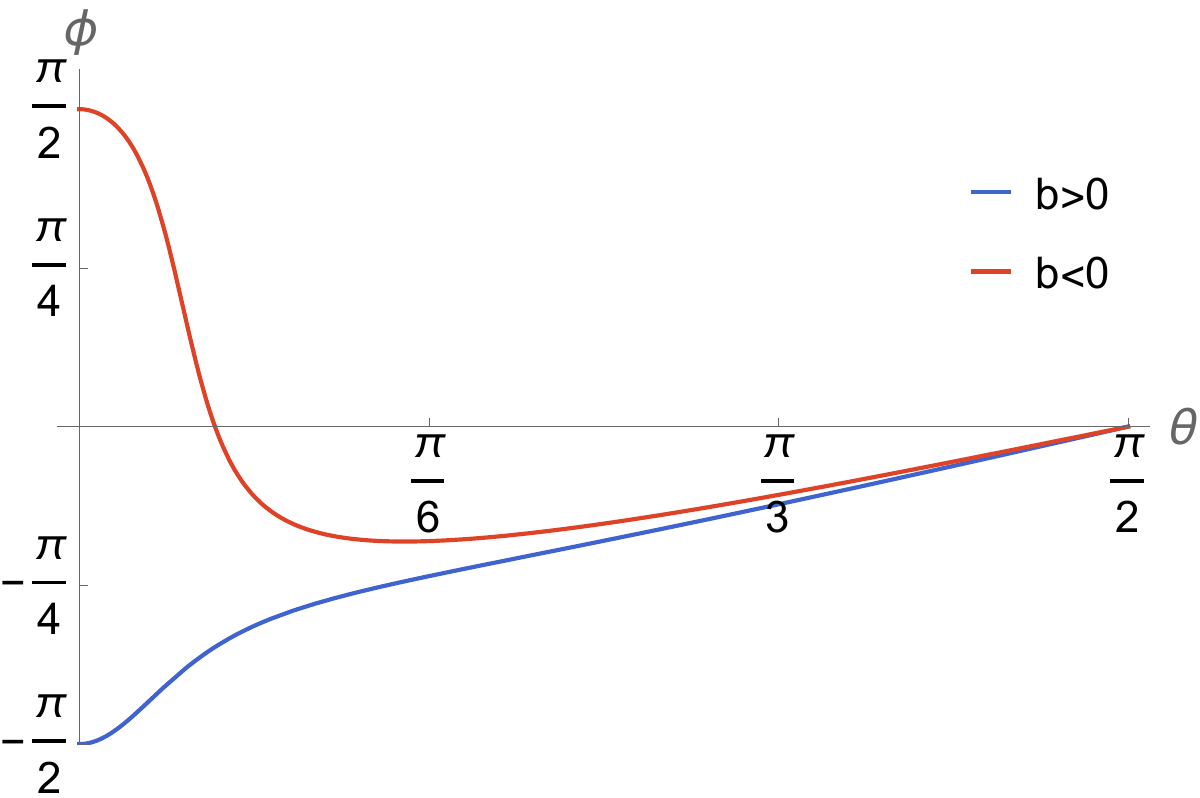}
    \caption{\justifying Dependence of the reflection phase shift $\phi=\arg(r)$ on incidence angle. It shows that, for small incidence angles, the phase shift strongly depends on the sign of the chirality parameter $b\propto\ell(J_1-J_2)$.}
    \label{fig:Reflection Phase Shift}
\end{figure}

As expected, in the normal-incidence limit $\theta\to 0$ the flexural wave is fully transmitted, while for $\theta\to\frac{\pi}{2}$ it is reflected with a $\frac{\pi}{2}$ phase shift. At intermediate angles, one has the frequency dependence shown in figure \ref{fig:Reflection Amplitude}. We notice that the small-frequency behavior depends on the sign of the chirality parameter $b\propto\ell (J_1-J_2)$, and is not monotonic for $b<0$. In particular, for $b<0$ the wave of frequency
\begin{eqnarray}
    \omega(\theta) = \frac{\gamma^2\ell^2(J_1-J_2)^2}{16\kappa_0^{1/2}\kappa_r^{3/2}}\frac{\sin^4\theta}{(1+\sin^2\theta)}\label{eq:omegatheta}
\end{eqnarray}
is fully reflected ($|r|=1$). As shown in figure \ref{fig:Reflection Phase Shift}, the phase shift of the reflected wave is also sensitive to the sign of $b$. In particular, the relative phase of the reflected wave at normal incidence $\theta\to 0$ is $\pm\frac{\pi}{2}$ for $b\lessgtr 0$, respectively. The dependence on sign of the chirality parameter $b\propto\ell(J_1-J_2)$, shown in figures \ref{fig:Reflection Amplitude} and \ref{fig:Reflection Phase Shift}, shows that the chirality of the superfluid is imparted on the reflection and transmission of flexural waves across the domain wall by the geometric coupling.

\section{Background extrinsic curvature}\label{sec:H0}

Let us consider now the effect of background curvature. Diagonalizing the curvature tensor at a point of the surface defines the principal curvatures $a$, $b$ (ie., $\frac{1}{a}$, $\frac{1}{b}$ are the principal curvature radii), and the local shape of the surface is given by the normal form $h_0(x,y)=\frac{1}{2}(ax^2+by^2)$ on the neighborhood of this point. An interesting special case is the uniaxial bending $b\to 0$ shown in figure \ref{fig:parabola}, since then the background Gaussian curvature $K_0=ab$ vanishes but the mean curvature $H_0=\frac{a+b}{2}$ is finite. Thus it corresponds to the case where the background has extrinsic curvature but no intrinsic curvature. We analyze this case first, leaving the effects of intrinsic curvature to the next section.

For the background shape $h_0=\frac{1}{2}ax^2$, the equilibrium conditions (\ref{eq:j0background},\ref{eq:h0background}) can be solved by vanishing background supercurrent and tension. Varying the action \eqref{eq:QuadraticAction} on this background, we find that the linearized equations of motion break translational symmetry. However, the functional form of the background is simple enough that the $x$-dependence can be absorbed into a phase redefinition,
    \begin{align}
        \delta \phi = \delta \theta +\ell H_0 x \partial _y\delta h,
    \end{align}
as we show in detail in Appendix \ref{app:parabolic}. The resulting coupled equations,
\begin{align}
\begin{pmatrix}
 \partial_t^2-u^2\Delta & 2\ell H_0 u^2\partial^2_{x,y}\\
  2\ell H_0 u^2\partial^2_{x,y} & \partial_t^2+\frac{\Delta^2}{4m_\kappa^2}- 4\ell^2H_0^2u^2\partial_y^2
\end{pmatrix}
\begin{pmatrix}
\delta \phi \\
\delta h
\end{pmatrix}
=0,\label{eq:WaveEquationParabolicBackground}
\end{align}
have translational symmetry and summarize the effect of extrinsic curvature on the propagation of the Goldstone and flexural modes. We see that, in a chiral superfluid $\ell\neq 0$, there is a coupling between the superfluid and geometry degrees of freedom, whose strength is set by the extrinsic curvature. For convenience, in equation \eqref{eq:WaveEquationParabolicBackground} we use the  dimensionless redefinitions of the curvature and the height field,
\begin{align}
    &H_0\mapsto \sqrt{\frac{\gamma_0}{\kappa_0}}H_0, &\delta h\mapsto \sqrt{\frac{\kappa_0}{\gamma_0}}\delta h.\label{rescalehH0}
\end{align}

\begin{figure}
    \centering
    \includegraphics[width=\linewidth]{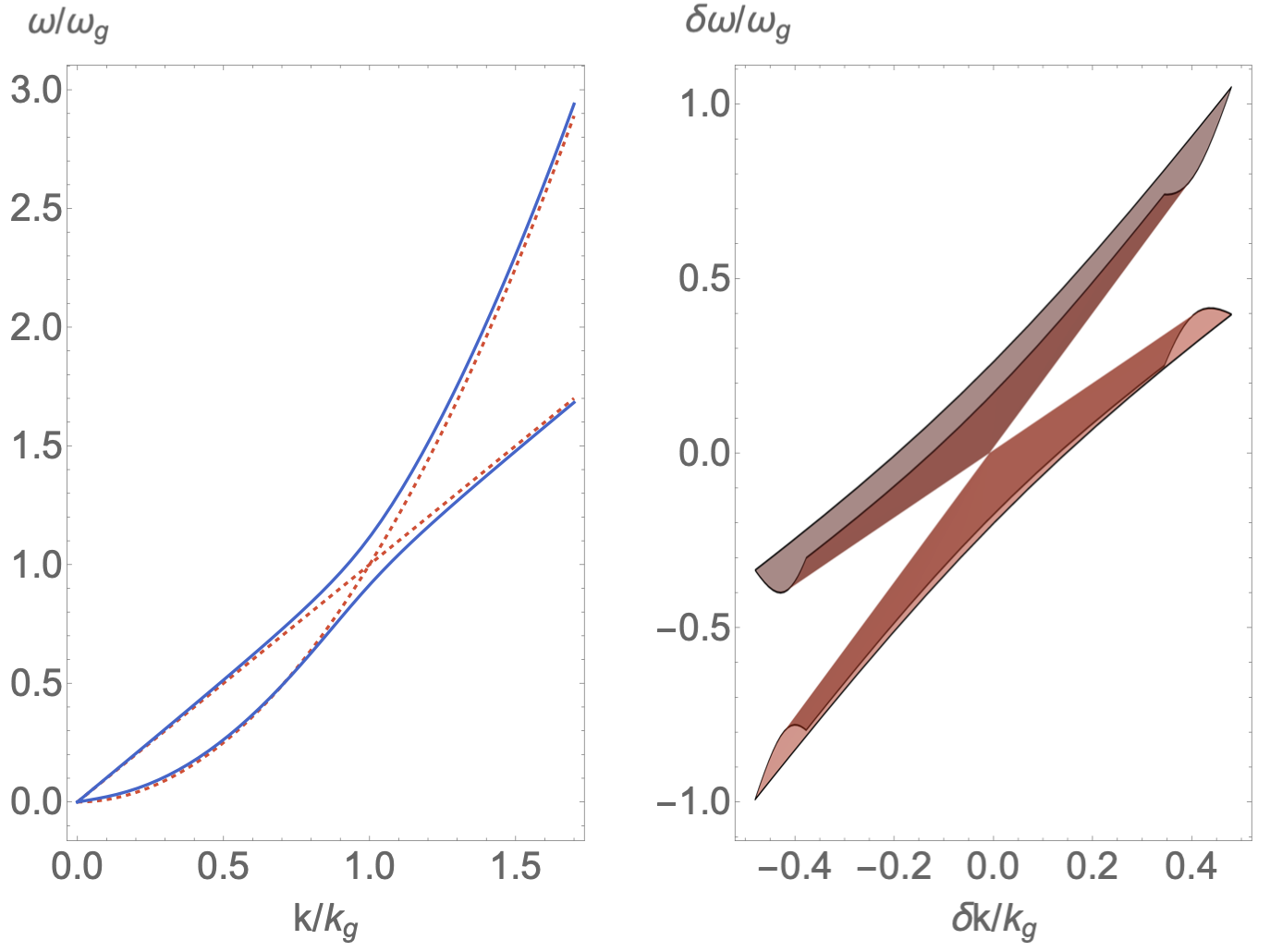}
    \caption{\justifying In a general direction $\theta=\pi/4$, the exterior curvature $H_0$ leads to a mixing of the flexural and Goldstone modes (left, blue), lifting the degeneracy of the bare spectrum at $k_g$ (left, red dashed). Still, for waves propagating along the principal axes, the gap closes at a point, leading to a tilted Dirac cone (right).}
    \label{fig:curvatureH0spectrum}
\end{figure}

The coupling leads to dressing of the flexural and the Goldstone modes. In fact, note that the bare bands (\ref{eq:Goldstoneomega}, \ref{eq:flexuralomega}) become degenerate at the momentum $k_g = 2m_{\kappa}u$, and frequency $\omega_g = 2m_\kappa u^2$. From \eqref{eq:WaveEquationParabolicBackground}, we obtain the normalized dispersion.
\begin{widetext}
\begin{align}
&\tilde{\omega}_{\pm}^2=\frac{\tilde{k}^2}{2}(\tilde{k}^2 +1+4\ell^2H_0^2\sin^2\theta) \pm \frac{\tilde{k}^2}{2}\sqrt{(\tilde{k}^2-1)^2+8\ell^2H_0^2\sin^2\theta(\tilde{k}^2+\cos2\theta )+16\ell^4H_0^4\sin^4\theta},
\label{eq:PreciseParabolicSpectrum}
\end{align}
\end{widetext}
where $\tilde{\omega}=\omega/\omega_g$, $\tilde{k}=k/k_g$, and again $\theta=\arctan\left(\frac{k_y}{k_x}\right)$ is the angle of propagation with respect to the curvature axis. Expanding around $k=k_g$, we find that the degeneracy is lifted,
\begin{equation}
    \omega_+(k_g)-\omega_-(k_g)=2\ell\omega_gH_0|\sin2\theta|+O(H_0^3).\label{eq:gapH0}
\end{equation}
The dispersion is shown in figure \ref{fig:curvatureH0spectrum} for $\theta=\pi/4$. Away from $k_g$, the dispersion is approximately given by the bare expressions, while close to $k_g$ one finds an avoided level crossing. This is similar to the formation of effective polaritonic modes \cite{huang1951lattice}, where here the effective degrees of freedom are combinations of the superflow and flexural waves.

Interestingly the gap \eqref{eq:gapH0} seems to vanish in the directions $\theta=\{0,\frac{\pi}{2},\pi,\frac{3\pi}{2}\},$ which correspond to the principal curvature axes. Indeed, in the direction $\theta=0,\pi$, which in our conventions is the axis where the principal curvature is $H_0$, the points $(k_x,k_y)=(\pm k_g,0)$ display a linear band touching. By expanding the fields around these points,
\begin{equation}
    \begin{pmatrix}
        \delta\phi\\
        \delta h
    \end{pmatrix}
    = e^{\pm i(\omega_g t \pm k_g x)}\begin{pmatrix}
        \delta\bar{\phi}\\
        \delta \bar{h}
    \end{pmatrix},
\end{equation}
we find that the slow fields satisfy a Dirac-type equations
\begin{equation}
    \pm(i\p_t \pm \mathcal{H}(k_x,k_y))
    \begin{pmatrix}
        \delta\bar{\phi}\\
        \delta \bar{h}
    \end{pmatrix}=0,
\end{equation}
with the effective Dirac Hamiltonian
\begin{equation}
    \mathcal{H}(k_x,k_y) = \frac{3u}{2}k_x\id_2-\ell H_0 u k_y\sigma^x -\frac{u}{2}k_x\sigma^z,
\end{equation}
which corresponds to a tilted Dirac cone (see figure  \ref{fig:curvatureH0spectrum}). Similarly, in the flat direction $\theta = \frac{\pi}{2},\frac{3\pi}{2}$, one has Dirac points at $(k_x,k_y)=(0,\pm k_g\sqrt{1-4\ell^2H_0^2})$, with the Dirac Hamiltonian

\begin{align}
    &\mathcal{H}(k_x,k_y) = \frac{3u}{2}u \left( 1-\frac{4}{3}\ell^2H_0^2\right)k_y\id_2 \\
    &-\ell H_0 u \sqrt{1-4\ell^2H_0^2}k_x\sigma^x-\frac{u}{2}\left( 1-4\ell^2H_0^2\right)^{\frac{3}{2}}k_y\sigma^z. \nonumber
\end{align}

Besides the closing of the gap along the principal directions, the full dispersion \eqref{eq:PreciseParabolicSpectrum} itself is anisotropic. In particular, we notice that for waves propagating in the curved principal direction, $\theta=0,\pi$, we recover the bare spectrum (\ref{eq:Goldstoneomega}, \ref{eq:flexuralomega}), while for waves propagating in the flat direction $\theta=\frac{\pi}{2},\frac{3\pi}{2}$, the frequency for the flexural mode is modified to
\begin{equation}
    \omega^2=\frac{k_y^4}{4m_\kappa^2}+4\ell^2 H_0^2 u^2k_y^2.\label{eq:modfiedflexuralH0y}
\end{equation}
At small momenta, $\omega\sim 2\ell H_0uk_y$ is dominated by a linear term, which corresponds to a membrane tension. Importantly, it dominates the dispersion at small wavevectors. In a general direction, the eigenfrequencies are mixed as in figure \ref{fig:curvatureH0spectrum}. Expanding in $H_0$, we obtain the modification to the small-momentum part of the dispersions as
\begin{align}
    \omega_+^2 &= \tilde{u}^2k^2\\
    \omega_-^2 &= \frac{k^4}{4\tilde{m}_\kappa^2} + 4\ell^2H_0^2u^2k^2\sin^4\theta,\label{eq:omegaminusH0}
\end{align}
where to leading order the shifts $\delta u = \tilde{u}-u$ and $\delta m_\kappa =\tilde{m}_\kappa-m_\kappa$ are given by
\begin{align}
    \frac{\delta u}{u}=\frac{\delta m_\kappa}{m_\kappa}=\frac{(\ell H_0\sin{2\theta})^2}{2}.
\end{align}
We see that \eqref{eq:omegaminusH0} contains a tension term which vanishes for waves propagating in the curved principal direction and is maximal for waves propagating in the flat principal direction. The anisotropic modification of the phase velocity of the Goldstone mode also leads to an anisotropic group velocity,
\begin{equation}
    \nabla_{\vec{k}}\omega_+ = u(\hat{r}+\ell^2H_0^2\sin{4\theta}\hat{\theta}).
\end{equation}

We see that, upon curving the background in one direction, the spectrum couples the Goldstone mode to the flexural mode, and its features are anisotropic. In particular, the modifications of the spectrum all vanish in the direction of the uniaxial bending. Geometrically, we notice that this direction is special because in case the full profile (background shape $h_0$ plus flexural mode $\delta h$) has no Gaussian curvature. We explore this point in more detail in the general profile where background Gaussian curvature is allowed.

\section{Background Gaussian curvature}\label{eq:K0}

We now consider the more general case when the background has not only mean curvature $H_0=\frac{a+b}{2}$ but also Gaussian curvature $K_0=ab$. Then the normal form of the background surface is given by $h_0(x,y) = \frac{1}{2}ax^2 + \frac{1}{2}by^2$, and the saddle-point equations for the background, (\ref{eq:j0background},\ref{eq:h0background}), imply a non-vanishing background supercurrent and tension
\begin{align}
    &\vec{J}_0 = \frac{\ell}{2}K_0 \,r\,\hat{\theta}, &\sigma_0 = -\frac{\gamma\ell^2}{2}K_0.
\end{align}
Again, we can absorb the coordinate dependence of the coefficients in the linearized equations by a phase redefinition
\begin{equation}
    \delta\phi = \delta\theta + \frac{\ell}{2}H_0(x\p_y-y\p_x)\delta h+\frac{\ell}{2}\cE_0(x\p_y+y\p_x)\delta h,
\end{equation}
where
\begin{eqnarray}
    \cE_0 = \frac{a-b}{2}
\end{eqnarray}
is the principal curvature anisotropy. It is not independent of $H_0$ and $K_0$, being related to them by
\begin{equation}
    H_0^2-\cE_0^2=K_0.
\end{equation}
The resulting equations coupling the superfluid phase and the flexural mode are
\begin{align}
\begin{pmatrix}
\partial _t^2-u^2\Delta
& 2\ell\cE_0 u^2\partial^2_{x,y}
\\
2\ell\cE_0 u^2\partial^2_{x,y}
& \partial_t^2 + \frac{\Delta^2}{4m_\kappa^2}-\ell^2(H_0^2+\cE_0^2)u^2\Delta\\
& \quad\quad\quad -2\ell^2H_0\cE_0u^2(\p_y^2-\p_x^2)
\end{pmatrix}
\begin{pmatrix}
\delta \phi\\
\delta h
\end{pmatrix}
= 0,\label{eq:equationsH0E0}
\end{align}
where we have again rescaled $\delta h$ and $H_0$ as in \eqref{rescalehH0}, and analogously for $\cE_0$. We give details on the derivation in Appendix \ref{app:tensionK}. Here we note that in the case where the Gaussian curvature vanishes, $\cE_0=H_0$, and we recover the equations in the previous section, \eqref{eq:WaveEquationParabolicBackground}.

An interesting special case of \eqref{eq:equationsH0E0} is that of maximal Gaussian curvature, $\cE_0=0, K_0=H_0^2$. Then the modes decouple, with phase fluctuations propagating with dispersion \eqref{eq:Goldstoneomega} and the flexural mode propagating with
\begin{align}
    \omega^2 = \frac{k^4}{2m_\kappa^2}+\ell^2K_0u^2k^2,\label{eq:tensionK0}
\end{align}
which now gives the flexural mode an isotropic tension so that $\omega\sim \ell\sqrt{K_0}uk$ at small momenta. In a general direction, there is a gap opening given by
\begin{align}
    \omega_+(k_g)-\omega_-(k_g) &=2\ell\omega_g|\cE_0||\sin2\theta|+O(\cE_0^3)\nonumber\\
    &\approx 2\ell\omega_g\sqrt{H_0^2-K_0}|\sin2\theta|.\label{eq:gapE0}
\end{align}
Thus in the more general case we find that the gap is actually proportional to the principal curvature anisotropy, and can be decreased by increasing the Gaussian curvature. Again the gap vanishes in the principal directions, on which we find tilted Dirac cones.

The modified dispersion gives, at small momenta,
\begin{align}
    \omega_+^2 &= \tilde{u}^2k^2,\\
    \omega_-^2 &= \frac{k^4}{4\tilde{m}_\kappa^2}+\frac{\ell^2}{2}[2H_0^2+\cE_0^2-4H_0\cE_0\cos{2\theta}\nonumber\\
    &\hspace{4.5cm}+\cE_0^2\cos{4\theta}]k^2,\label{eq:tensiongeneral}
\end{align}
where, to the lowest order,
\begin{align}
    \frac{\delta u}{u}=\frac{\delta m_\kappa}{m_\kappa}=\frac{(\ell \cE_0\sin{2\theta})^2}{2},
\end{align}
and the modified group velocity for the Goldstone mode becomes
\begin{equation}
    \nabla_{\vec{k}}\omega_+ = u(\hat{r}+\ell^2\cE_0^2\sin{4\theta}\hat{\theta}).
\end{equation}

\begin{figure}[]
    \centering
    \includegraphics[width=.9\linewidth]{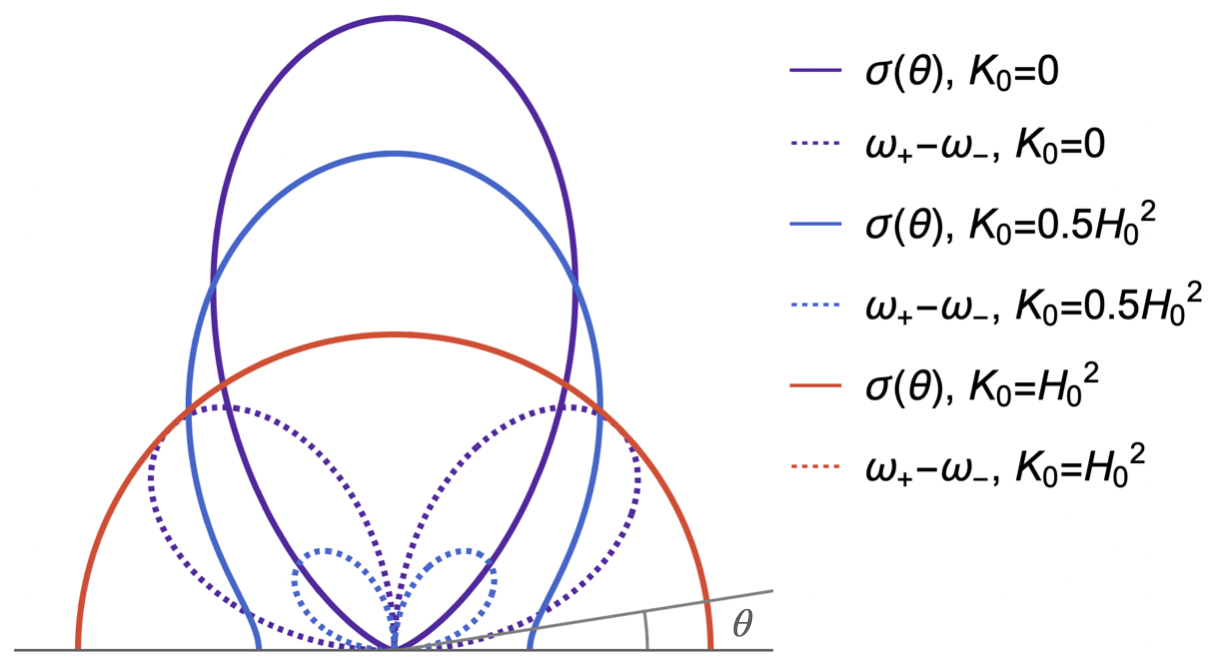}
    \caption{\justifying Angular dependence of the effective tension, $r = \sigma(\theta)$ and gap $r = (\omega_+-\omega_-)|_{k_g}(\theta)$ for different values of Gaussian curvature, and fixed mean curvature. The gap is maximal for vanishing Gaussian curvature, in which case the effective tension on the spectrum of flexural waves is maximal on the direction transverse to the curved direction. As the Gaussian curvature increases, the distribution of tension becomes isotropic and the gap vanishes.}
    \label{fig:tensiongaptheta}
\end{figure}

Note that equation \eqref{eq:tensiongeneral} implies a more general anisotropic effective tension in the propagation of flexural waves. It reduces to the tension \eqref{eq:modfiedflexuralH0y} in the direction transverse to the bending direction in the limit of no Gaussian curvature $K_0\to 0$ and to \eqref{eq:tensionK0} in the isotropic limit $K_0 \to H_0^2$. For intermediate values, the tension is anisotropic, being stronger in the direction transverse to the principal direction of stronger curvature. Conversely, the gap always vanishes on the principal directions, and its magnitude is maximal in the uniaxial bending limit of no Gaussian curvature and vanishes in the isotropic limit. We summarize the angular dependence of this parameters in figure \ref{fig:tensiongaptheta}.

\section{Conclusion}\label{sec:conclusion}

We studied the effects that the coupling between a chiral superfluid and the background geometry has on the propagation of the flexural deformations and of the superfluid phase, or Goldstone mode. We considered this problem by linearizing the full action around a general background, and solving the resulting equations of motion in different cases.

In the case of a background supercurrent, we found that the leading effect is a modification of the boundary conditions for the flexural mode. If the superfluid is localized to a finite region of the substrate, then it might drag the flexural mode along the boundary, leading to boundary flexural waves. The condition for this effect to appear is chiral: only if the supercurrent flow along the boundary points on the positive direction with respect to the cross product of the condensate angular momentum and the normal direction to the boundary. Additionally, the presence of boundary modes changes the reflection and transmission coefficients for flexural waves across the boundary, which also become dependent on the chirality of the superfluid through the direction of the background supercurrent.

In the case of a curved background, we found that anisotropy between the principal curvatures leads to the dressing of the Goldstone and flexural modes, with a gap opening for the effective degrees of freedom. The exception is for waves propagating along the principal directions of the curvature tensor, for which we found that the gap closes forming a tilted Dirac point. In the small-momentum expansion of the dispersion relations, we find that the curvature anisotropy gives anisotropic corrections to the superfluid phase and group velocities, as well as to the mass stiffness of the flexural waves. Moreover, we found that bending the background along one direction generates a tension for the flexural waves propagating in the transverse direction, and the tension becomes isotropic in the limit of maximal Gaussian curvature $K_0=H_0^2$.

Our work reveals dynamical signatures of the geometric response of chiral condensates. Interestingly, the formation of a chiral condensate modifies the propagation of the flexural mode. The appearance of a tension in the case of a curved background can significantly change the amplitude of shape fluctuations at large scales, with possible implications even for the thermal properties of the substrate shape \cite{shankar2021thermalized}. Finally, exploring how our findings generalize to the charged case of chiral superconductors is an interesting future direction.

\section{Acknowledgments}

We thank Michael Stone for insightful discussions. This work was supported by National Natural Science Foundation of China (NSFC) under Grant No. 23Z031504628 (G.C., R.M., and Q.-D.J.), Jiaoda2030 Program Grant No.WH510363001, TDLI starting up grant, and Innovation Program for Quantum Science and Technology Grant No.2021ZD0301900 (Q.-D.J.).

\bibliography{main}

\appendix

\section{Boundary Mode and Anomalous Reflection}\label{Derivation for 1st Setup}

In our setup, the boundary condition is given by
\begin{align}
\partial_x^3 \delta h|_{+}-\partial_x^3 \delta h|_{-}=b k_y^2\delta h,
\end{align}
in terms of the Fourier transform of $\delta h$ in the $y$ direction (we will not use an alternative notation such as $\mathcal{F}[\delta h](k_y)$ for the Fourier transform since the meaning is clear from the context). Here, we use the notation $b=\frac{\gamma \ell u^2 (J_1-J_2)}{\kappa_r}$. The bulk equation of motion,
\begin{equation}
    [\kappa\p_t^2+\kappa_r(\p_x^2-k_y^2))^2]\delta h=0,
\end{equation}
can be solved by $\delta h(x,t) = e^{i(\omega t-k_xx)}$, where
\begin{eqnarray}
    \omega^2=a^2(k_x^2+k_y^2)^2,\label{eq:omegakxkyboundary}
\end{eqnarray}
and we defined the notation $a = \sqrt{\frac{\kappa_r}{\kappa}}$. Note that this equation is of fourth order in $k_x$, which can be real or imaginary. We consider first the band for which
\begin{eqnarray}
    |\omega|<ak_y^2.
\end{eqnarray}
It leads to four exponential solutions $e^{i\omega t+\lambda x}$, with
\begin{align}
    &\lambda = \pm\lambda_{1,2}, &\lambda_{1,2}=\sqrt{k_y^2\pm\frac{\omega}{a}}.\label{eq:lambdaboundary}
\end{align}
This band is given by a one-dimensional dispersion relation
\begin{equation}
    \omega(k_y)=\pm a(\lambda^2-k_y^2),\label{eq:omegaboundary}
\end{equation}
where the $\lambda\in\{\pm\lambda_{1,2}\}$ are determined by the boundary conditions. In our domain wall geometry (figure \ref{fig:backgroundsupercurrent}), we must select exponentially localized solutions
\begin{align}
\delta h(x)=
\left\{\begin{matrix}
C_1 e^{-\lambda _1 x}+C_2e^{-\lambda _2 x}, &\quad x>0 \\
C_3 e^{\lambda _1 x}+C_4e^{\lambda _2 x }, &\quad x<0
\end{matrix}\right.,
\end{align}
and the boundary conditions at $x=0$ become
\begin{align}
&C_1+C_2= C_3+C_4 ,\\
&\lambda_1 C_1 + \lambda_2 C_2 = -\lambda_1 C_3 - \lambda_2 C_4 ,\\
&\lambda_1^2 C_1 + \lambda_2^2 C_2 = \lambda_1^2 C_3 + \lambda_2^2 C_4 ,\\
&(\lambda_1^3+bk_y^2)C_1 + (\lambda_2^3+bk_y^2)C_2=-\lambda_1^3 C_3 - \lambda_2^3C_4.
\end{align}

\begin{figure}
    \centering
    \includegraphics[width=1.0\linewidth]{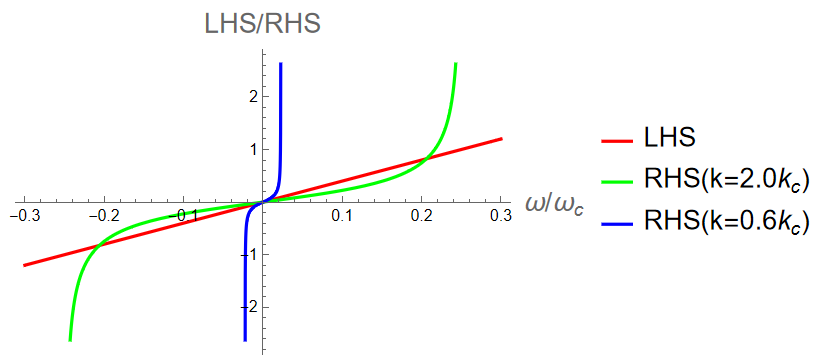}
    \caption{\raggedright Here the graphic solution of the equation \ref{Determinant} under different $k$s.For small $k$ value , the slope at $k=0$ is large, thus there is only one intersection point at zero. Non zero solutions are attained when one has small enough slope at 0, which requires $k > k_c$. }
    \label{fig:graphicsol}
\end{figure}

Nontrivial solutions can only appear for vanishing determinant, which is given by
\begin{equation}
    2(\lambda_1-\lambda_2)^2(\lambda_1+\lambda_2)(bk_y^2-2\lambda_1\lambda_2(\lambda_1+\lambda_2)).
\end{equation}
The first root, $\lambda_1=\lambda_2$, leads only the trivial solution $C_1 = C_2 = C_3 = C_4 = 0$. The second root $\lambda_1=-\lambda_2$ cannot be satisfied since the sign of $\lambda_{1,2}>0$ is fixed by the boundary conditions at infinity. For the third root, note that
\begin{align}
bk_y^2(\lambda_1 - \lambda_2)&=2\lambda_1 \lambda_2 (\lambda_1 + \lambda _2)(\lambda_1 - \lambda_2) \\
&= \frac{4}{a}\omega \lambda _1\lambda _2 \\
\Rightarrow\frac{4\omega }{abk_y^2} = \frac{1}{\lambda _2}&-\frac{1}{\lambda _1},
\end{align}
which leaves the equation
\begin{align}
\frac{4\omega}{abk_y^2} = \frac{1}{\sqrt{k_y^2-\frac{\omega}{a}}}-\frac{1}{\sqrt{k_y^2+\frac{\omega}{a}}}.\label{Determinant} 
\end{align}
for the dispersion $\omega(k_y)$ of the boundary waves. The solution corresponds to the expression \eqref{eq:omegaboundary} when the localization length is determined by the boundary conditions. We illustrate the existence of this solution graphically in figure \ref{fig:graphicsol}. The left-hand-side (LHS) and the right-hand-side (RHS) of \eqref{Determinant} are plotted as functions of $\omega$, with the other parameters fixed, so that solutions are given by the intersections of these curves. For $b<0$, there is no intersection except for the trivial one at $k_y=0$, so that there are only propagating boundary waves for $b>0$, ie.,
\begin{align}
\left\{\begin{matrix}
J_1>J_2, &\quad \text{if }\ell>0 \\
J_1<J_2, &\quad \text{if }\ell<0
\end{matrix}\right..\label{eq:J12ell}
\end{align}
Furthermore, the extra intersections at $\omega>0$ only appear if the slope of the LHS (the straight line) is larger than the slope of the RHS at the origin, which gives the condition
\begin{align}
    \frac{4}{abk_y^2}>\frac{1}{a|k_y|^3}\Rightarrow |k_y|>\frac{b}{4},
\end{align}
so that the momentum along the boundary has a minimal value of
\begin{eqnarray}
    k_c = \frac{\gamma \ell u^2 (J_1-J_2)}{4\kappa_r}.\label{eq:kcJ12}
\end{eqnarray}

For $k_y$ near $k_c$, $\omega$ is small, and one can determine the approximate dispersion by expanding equation \eqref{Determinant}. This gives
\begin{align}
&\frac{4 \omega }{abk_y^2} = \frac{\omega }{a|k_y|^3}+\frac{5}{8}\frac{\omega ^3}{a^3|k_y|^7}+O (\omega^4),\\
&\Rightarrow \omega (k_y) \approx \pm 2 a\sqrt{\frac{2}{5}} k_y^2\sqrt{\frac{|k_y|}{k_c}-1},
\end{align}
which agrees with the numerical solution shown in figure \ref{fig:boundaryspectrum}.

Consider now the band corresponding to
\begin{eqnarray}
    |\omega|>ak_y^2.
\end{eqnarray}
Then \eqref{eq:omegakxkyboundary} has two propagating solutions $e^{i(\omega t- k_x x)}$, with
\begin{align}
    k_x = \pm\sqrt{\frac{|\omega|}{a}-k_y^2},
\end{align}
and two exponential solutions $e^{i\omega t+\lambda x}$, with
\begin{align}
    &\lambda = \pm\sqrt{\frac{|\omega|}{a}+k_y^2}.
\end{align}
These solutions play a role in the scattering of the flexural mode across the boundary.

Specifically, we find that a superposition of the incident, reflected and transmitted waves is not enough to solve all the boundary conditions. Instead, one needs to include a component of the localized solutions,
\begin{align}
\delta h(x)=
\left\{\begin{matrix}
e^{ik_xx}+re^{-ik_xx}+C_1e^{\lambda x}, &\quad x<0 \\
te^{ik_xx}+C_2e^{-\lambda x}, &\quad x>0
\end{matrix}\right..
\end{align}
The boundary conditions become
\begin{align}
    &-r-C_1+t+C_2 = 1,\\
     &ik_xr-\lambda C_1 + ik_xt - \lambda C_2 = ik_x,\\
    &k_x^2r - \lambda ^2 C_1 - k_x^2t + \lambda ^2C_2 = -k_x^2,\\
    &ik_x^3r+(bk_y^2+ik_x^3)t+\lambda ^3C_1 + (\lambda ^3+bk_y^2)C_2 = ik_x^3.
\end{align}
The solution is given by
\begin{align}
    r &= \frac{-b \sin^2 \theta  \sqrt{1+\sin^2\theta} }{4i\sqrt{\frac{\omega}{a}} \sqrt{1-\sin^4\theta }+b\sin^2\theta (\sqrt{1+\sin^2\theta }+i \cos\theta  ) },\nonumber\\
    t &= \frac{ \cos\theta (b\sin^2\theta +4\sqrt{\frac{\omega}{a} }\sqrt{1+\sin^2\theta }  ) }{4\sqrt{\frac{\omega}{a}} \sqrt{1-\sin^4\theta }+b\sin^2\theta (\cos\theta  -i\sqrt{1+\sin^2\theta }   )},\nonumber
\end{align}

where $\theta=\arctan\left(\frac{k_y}{k_x}\right)$ is the incidence angle. Importantly, the amplitudes of the localized solutions, $C_{1,2}$, are non-vanishing, and play a role in determining the reflection and transmission coefficients. Note also that these expressions give $(r,t)\to (0,1)$ in the limit $b\to 0$, corresponding to full transmission. For $b\neq 0$, we can define the phase shift $\phi$ of the reflected wave by $r = \lvert r \rvert e^{i\phi}$. Then we find that
\begin{align}
\tan \phi = \frac{4  \sqrt{1-\sin^4\theta }+b_c\sin^2\theta \cos\theta   }{b_c\sin^2\theta \sqrt{1+\sin^2\theta }  } .
\end{align}
Here, the definition of $b_c \equiv b \sqrt{\frac{a}{\omega}}$.In particular, in the limit of normal incidence $\theta\to 0$, $\tan \phi \rightarrow +\infty$, so that there is a $\frac{\pi}{2}$ phase shift between the incident and the reflected wave.

\section{Background with extrinsic curvature}\label{app:parabolic}

At the solution $h_0=\frac{1}{2}ax^2$, $J_0 = 0$ and $\sigma = 0$ of the background equilibrium conditions (\ref{eq:j0background},\ref{eq:h0background}) we find, by varying the action \eqref{eq:QuadraticAction}, the linearized equations of motion
    \begin{align}
        &\partial_\mu\partial^\mu\delta \theta +\frac{\ell}{2}ax \partial^{\mu}\partial_{\mu}\partial _y\delta h =0, \\
        &\kappa _0 \partial_t^2 \delta h + \kappa_r \Delta^2 \delta h=\frac{\gamma_0 \ell}{2}a\partial_{\mu}(x\partial ^{\mu}\partial _y\delta \theta )-\frac{\gamma_0 \ell}{2}au^2\partial_x\partial _y\delta \theta \nonumber \\
        &\hspace{1.5cm}+\frac{\gamma_0 \ell^2}{4}a^2\partial ^{\mu}(x^2\partial _{\mu}\partial _y^2\delta h)+\frac{\gamma_0 \ell^2}{4}a^2u^2\partial_x(x\partial _y^2\delta h) \nonumber \\
        &\hspace{1.5cm}-\frac{\gamma_0 \ell^2}{4}a^2u^2x\partial _y^2\partial _x\delta h + \frac{\gamma_0 \ell^2}{4}a^2u^2\partial _y^2\delta h.       
    \end{align}
We note that, by the phase redefinition
    \begin{align}
        \delta \phi = \delta \theta +\frac{\ell}{2}ax \partial _y\delta h,
    \end{align}
the first equation can be rewritten as
    \begin{align}
        \partial _{\mu}\partial ^{\mu}\delta \phi + \ell au^2\partial_x \partial_y \delta h = 0.
    \end{align}
Likewise, one can use this definition to bring the second equation to the form

\begin{align}
&\kappa _0 \partial _t^2 \delta h + \kappa _r \Delta^2 \delta h = \frac{\gamma_0 \ell}{2}a\partial _{\mu}(x \partial ^{\mu}\partial _y \delta \phi)-\frac{\gamma_0 \ell}{2}au^2\partial _x\partial _y\delta \phi \nonumber \\
  &\hspace{1.5cm}+\frac{\gamma_0 \ell^2}{2}a^2u^2\partial _y^2\delta h + \frac{\gamma_0 \ell^2}{2}a^2u^2\partial _x(x\partial_y^2\delta h)\label{eq:fullsecondeq}
\end{align}
Using the first equation of motion, one can show that
\begin{align}
\partial _{\mu}(x\partial ^{\mu}\partial _y \delta \phi) =-\ell au^2x\partial _x\partial _y^2\delta h-u^2\partial _x\partial _y\delta \phi
\end{align}

and, finally, substituting in \eqref{eq:fullsecondeq} reduces the second equation to
\begin{align}
\kappa _0\partial _t^2\delta h+\kappa _r\Delta^2\delta h=\gamma_0 \ell^2a^2u^2\partial 
_y^2\delta h -\gamma_0 \ell au^2\partial _x\partial _y\delta \phi.
\end{align}


\section{Background with Gaussian curvature}\label{app:tensionK}
\subsection{Background Stress $\sigma$}

A general quadratic background with nonzero Gaussian curvature would spontaneously generate current. Here we demonstrate that one needs to apply external force to balance the current. Moreover, the tension we need to apply is proportional to Gaussian curvature. 
We recall that the current $\vec{J}=(\nabla \theta + \ell \vec{\Omega})$. Here, $\vec{\Omega}$ is the spin connection. In \cite{jiang2022geometric}, it is demonstrated that under the symmetric gauge $\hat{e}_1^2 = \hat{e}_2^1$ the spin connection to the leading order in gradient expansion is
\begin{align}
\Omega_i=\frac{\ell}{2}\epsilon ^{jk}\partial _k(\partial _jh\partial _ih) .
\end{align}
For the background $h(x,y) = \frac{1}{2}(ax^2 + by^2)$, the spin connection is
\begin{align}
\vec{\Omega} = \frac{ab}{2}r \hat{e}_{\theta}.
\end{align}
We notice that $\nabla \times \vec{\Omega} \ne 0$. This means one can not eliminate the spin connection via a gauge transformation $\theta \to \theta + \ell \chi$, $\vec{\Omega} \to \vec{\Omega}-\nabla \chi$. The background has nonzero current,
\begin{align}
\vec{J}_0=\frac{ \ell}{2}abr\hat{e}_{\theta }.  
\end{align}

Due to the non-zero background current, external stress is necessary to stabilize the configuration. The external stress we need to apply could be determined from the equations \eqref{eq:j0background} and \eqref{eq:h0background},

\begin{eqnarray}
\sigma (a+b)=-\frac{\gamma_0 \ell^2}{2}ab(a+b),
\end{eqnarray}
so that $\sigma = -\frac{\gamma \ell^2}{2}K_0$. 

\subsection{Equations of motion}

We first start with the equation of motion for superfluid under the background $h(x,y)=\frac{1}{2}ax^2+\frac{1}{2}by^2$,
\begin{align}
&\partial _{\mu}\partial ^{\mu}\delta \theta +\frac{\ell}{2}\partial ^{\mu}(ax\partial _y\partial _{\mu}\delta h-by\partial _x \partial _{\mu} \delta h)+ \\ \nonumber
&\hspace{3cm}+\frac{\ell}{2}u^2(a-b)\partial _x \partial_y \delta h =0.  
\end{align}
For the second term, we have
\begin{align}
 \partial ^{\mu}(ax\partial _y\partial _{\mu}\delta h)&=\partial ^{\mu}\partial _{\mu}(ax\partial _y\delta h)+au^2\partial _x\partial _y\delta h\\  
\partial ^{\mu}(by\partial _x\partial _{\mu}\delta h)&=\partial ^{\mu}\partial _{\mu}(by\partial _x\delta h) +bu^2\partial _x\partial _y\delta h.
\end{align}
We can simplify the equation of motion by a redefinition of phase $\delta \theta$,
\begin{align}
\delta \phi = \delta \theta +\frac{\ell}{2}(ax\partial _y\delta h-by\partial _x\delta h). 
\end{align}
The first equation is thus rewritten in a translational-invariant way,
\begin{align}
\partial _{\mu}\partial ^{\mu}\delta \phi+\ell u^2(a-b)\partial _x\partial _y\delta h=0. \label{eq:CurrentQuadraticBackground}
\end{align}
The second linearized equation is given by
\begin{align}
&0=-\kappa _0\partial _t^2\delta h+\gamma_0 \ell^2u^2ab\Delta\delta h-\kappa _r\Delta ^2\delta h \nonumber\\
&+\frac{\gamma_0 \ell}{2}u^2(b-a)\partial _x\partial _y\delta \theta +\frac{\gamma_0 \ell}{2}\partial _{\mu}(ax\partial _y \partial ^{\mu}\delta \theta -by\partial _x\partial ^{\mu}\delta \theta ) \nonumber\\
&+\frac{\gamma_0 \ell^2}{4}[\partial _x \partial ^{\mu}(b^2y^2\partial _x \partial _{\mu}\delta h)+\partial _{y}\partial ^{\mu}(a^2x^2\partial _{y}\partial _{\mu}\delta h)\nonumber \\
&-\partial _x\partial ^{\mu}(abxy\partial _{y}\partial _{\mu}\delta h)-\partial _y\partial ^{\mu}(abxy\partial _x \partial_{\mu}\delta h)] \nonumber  \\
&+\frac{\gamma_0 \ell^2 u^2}{2} \left( a^2\partial_y^2 + b^2\partial_x^2 + \frac{1}{2}ab \Delta \right)\delta h.
\end{align}
We can use a trick similar to the case of external curvature above to simplify the second equation. After replacing $\delta \theta$ with $\delta \phi$ and using equation \eqref{eq:CurrentQuadraticBackground}, one finds
\begin{align}
&-\kappa _0\partial _t^2\delta h+\gamma_0 \ell^2u^2ab\Delta\delta h-\kappa _r\Delta ^2\delta h \nonumber \\
&+\gamma_0 \ell u^2(b-a)\partial _x \partial _y\delta j + \gamma_0 \ell^2u^2(a^2\partial_y^2+b^2\partial _x^2)\delta h=0.\nonumber
\end{align}

 \clearpage 

 \newpage

\end{document}